\documentclass[aps,pre,onecolumn,superscriptaddress,groupedaddress,showpacs]{revtex4}

\usepackage{graphicx}
\usepackage{amsfonts}

\begin{document}

\title[Star copolymers in porous environments]
{Star copolymers in porous environments: scaling and its manifestations}
\author{V.  Blavatska}
\email[]{viktoria@icmp.lviv.ua}
\affiliation{Institute for Condensed
Matter Physics of the National Academy of Sciences of Ukraine,\\
79011 Lviv, Ukraine}
\author{C. von Ferber}
\affiliation{Applied Mathematics Research Centre,
Coventry University, Coventry, UK}
\affiliation{Institute of Physics, Universit\"at Freiburg,
                D-79104 Freiburg, Germany}
\author{Yu.  Holovatch}
\affiliation{Institute for Condensed
Matter Physics of the National Academy of Sciences of Ukraine,\\
79011 Lviv, Ukraine}
\affiliation{Institut f\"ur Theoretische Physik, Johannes Kepler
Universit\"at Linz, A-4040, Linz, Austria}

\begin{abstract}

We consider star polymers, consisting of two different polymer species, in a
solvent subject to  quenched correlated structural
obstacles. We assume that the disorder is correlated with a
power-law decay of the pair correlation function $g(x)\sim
x^{-a}$. Applying the field-theoretical renormalization group approach in $d$ dimensions,
we analyze different scenarios of scaling behavior working to first
order of a double $\varepsilon=4-d$, $\delta=4-a$ expansion. We discuss the
influence of the correlated disorder on the resulting scaling laws
and possible manifestations such as  diffusion controlled reactions in the vicinity
of absorbing traps placed on polymers  as well as the
 effective short-distance
interaction between star copolymers.

\end{abstract}

\maketitle

\section{Introduction}

Understanding the behavior of polymer macromolecules in solutions in the
presence of structural obstacles is of great interest
in polymer physics. The presence of defects often leads
 to a large spatial inhomogeneity and may create pore spaces of fractal structure
 \cite{Dullen79}.
 Such situations can be encountered in studying, e.g., polymer diffusion through
 microporous membranes
 \cite{Cannel80} or within colloidal solutions \cite{Pusey86}.

Solutions of polymer macromolecules in disordered environment are subject to
intensive studies. Numerous simulations
\cite{Kremer81,Lee88,Grassberger93,Meir89,Rintoul94,Ordemann00}
and analytical studies
\cite{Meir89,Sahimi84,Rammal84,Kim87,Chakrabarti81,Kim83,Harris83,Roy82}
have focussed on the case of uncorrelated structural defects at the
percolation threshold of the remaining accessible sites, this situation is shown to
  alter significantly
the universal behavior of polymer macromolecules.
Recently, another special type of disorder has been brought to attention,
which display correlations in mesoscopic scale. This case can be described within the
frames of a model with
long-range-correlated quenched defects, considered in Refs. \cite{Weinrib83,Prudnikov,Korutcheva}
in the context of magnetic phase transitions. Here, structural defects are characterized by
a pair correlation function $g(x)$, which in $d$ dimensions falls off at large distance $x$ according to a 
power law:
\begin{equation} g(x)\sim x^{-a}.\label{parcor}
\end{equation}
In general, any value of $0\leq a \leq d$ can be realized by defects
that form clusters of fractal dimension $d_f=d-a$. For integer dimension $d_f$
these include the following special cases: uncorrelated point-like defects ($d_f=0$),
mutually uncorrelated straight lines of random orientation ($d_f=1$), mutually uncorrelated
planes of random orientation ($d_f=2$).
The influence of such long-range correlated defects on
the universal properties of single polymer has been  analyzed within the renormalization
group approach in Refs. \cite{Blavatska01,Blavatska01a}.

To describe the universal properties of polymer chains in good solvents, one may due to universality in the long chain limit consider the model of self-avoiding walks (SAWs) on a regular lattice
\cite{Cloizeaux,Gennes}.
In particular, the average square end-to-end distance $\langle R_e^2\rangle$
 and the number of
configurations $Z_N$  of SAWs with $N$ steps obey in the asymptotic limit $N\to\infty$
the following scaling laws:
\begin{equation}\label{scaling}
 \langle R_e^2 \rangle
\sim N^{2\nu},\mbox{\hspace{3em}}\mbox{\hspace{3em}}
Z_N \sim R^{2-\eta-1/\nu}.
\end{equation}
Here, the second equation shows the power law in terms of the effective polymer size
$R\equiv \sqrt{\langle R_e^2\rangle}\sim N^{\nu}$,  $\nu$ and $\eta$ are universal
exponents that only depend on the
space dimensionality $d$.  For $d=3$, high order renormalization group estimates are \cite{Guida98}
$\nu=0.5882\pm 0.0011$ and $\eta=0.0284\pm0.0025$.

The  theory  can be generalized to describe star polymers, which consist of $f$
linear polymer chains or SAWs, linked together at their end-points. The study of star
polymers is of great
interest since they serve as building blocks of polymer networks
\cite{Duplantier-1989I,Schafer-1992} and
can be used to model complex polymer micellar systems and gels
 \cite{Grest96,Likos01,Ferber02}.
For a single star with $f$ arms of $N$ steps (monomers) each,
the number of possible configurations obeys a power law in terms of the size $R$
of the isolated chain of $N$ monomers \cite{Duplantier-1989I,Schafer-1992}:
\begin{equation}
\label{gamstar}
Z_{N,f} \sim R^
{\eta_f-f\eta_2}.
\end{equation}
Here, the exponents
$\eta_f$ are universal star exponents, depending on the
number of arms $f$ ($\eta_1=0$, $\eta_2=1/\nu-2-\eta$).
Scaling properties of star polymers are well studied both numerically
\cite{Grest87,Batoulis89,Ohno94,Shida96,Barrett87,Hsu04} and
analytically \cite{Schafer-1992,Miyake83,Duplantier86,Ohno88,Ohno89,Ohno91,Ferber95,Ferber96}.
It has been shown that the presence of
 long-range-correlated disorder
may have interesting consequences for the scaling properties of polymer stars such as entropic separation of polymers
according to their architecture  \cite{Blavatska06}.

\begin{figure}[b!]
\caption{\label{Figure1} (color online) Schematic representation of copolymer stars consisting of two polymer
species (denoted as red and blue). Solid lines present species behaving like SAWs, dashed
lines present RWs. The two different sets in each example may further be
either mutually avoiding or mutually ``transparent".}
 \begin{center}
\includegraphics[width=12cm]{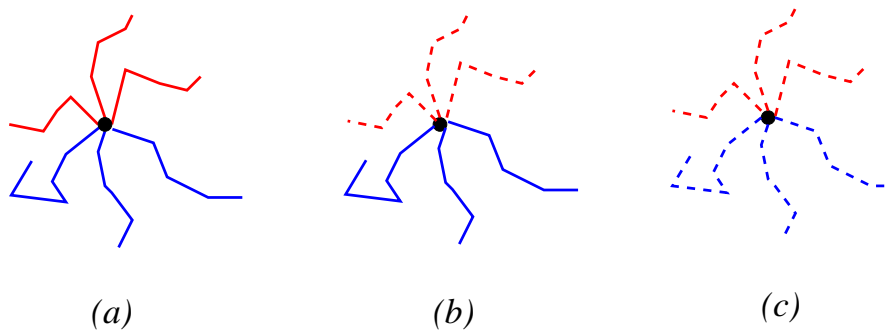}
\end{center}
 \end{figure}

Linking together polymers of different species, we receive
non-homogeneous  star polymers with a much richer scaling behavior
\cite{Schafer91,Ferber97,Ferber97e,Ferber99}.
A particular case is  the star copolymer, consisting of polymer chains of two different species.
It has been  shown \cite{Ferber97}, that the number of configurations $Z_{f_1f_2}$ of
a copolymer star with $f_1$ arms of species 1 and $f_2$ arms of species 2
scales as:
\begin{equation}\label{comstar}
Z_{f_1f_2} \sim(R)^
{\eta_{f_1f_2}-f_1\eta_{2 0}-f_2\eta_{0 2}},
\end{equation}
where $\eta_{f_1f_2}$ constitutes a family of copolymer star exponents.  These exponents
 are universal and depend only on space dimension $d$ and the number of chains $f_1$, $f_2$,
 as well as three different types of fixed points that govern the rich scaling behavior
 \cite{Schafer91}.

Depending on the temperature,
a situation may occur, where one or more of the inter- or intrachain interactions vanishes.
Indeed, for each polymer system one finds a so-called $\Theta$-temperature, at which
attractive and repulsive interactions between monomers compensate each other (see, e.g.,
\cite{Cloizeaux,Gennes}).
Such a polymer chain can effectively be described by a simple random walk (RW). In this case,
scaling laws (Eq. (\ref{scaling})) hold with exponents: $\nu=1/2$, $\eta=0$.
As a result, for example, there may be only mutual excluded-volume interactions between chains
of different species, while chains of the same species can freely intersect.
That is, some species behave effectively like RWs.
Within a copolymer star, the subset of chains of such species builds up a substar of
random walks, possibly
avoiding the second part of the star,
which can be either of random walks or self-avoiding random walks (see Fig. \ref{Figure1}).
Cates and Witten \cite{Cates87} have shown, that this situation can also be interpreted
as describing the  absorption of diffusive particles on polymers.

Another example, where star exponents govern physical behavior concerns the short-range
interaction  between cores  of star polymers in a good solvent
 \cite{Duplantier-1989I,Ferber97c,Ferber00}.  The mean force $F_{fg}(r)$ acting
on the centers of two stars with functionalities $f$ and $g$ is inversely proportional
to the distance $r$ between their cores:
\begin{equation}
F_{fg}(r)=k_BT\frac{\Theta_{fg}}{r},
\end{equation}
with $k_BT$ denoting the thermal energy, $\Theta_{fg}$ the universal contact exponent,
 related to the family of exponents
of star polymers by scaling relations:
\begin{eqnarray}
\Theta_{fg} =\eta_{f} + \eta_{g}- \eta_{f+g} \label{s}.
\end{eqnarray}
We are interested to generalize this relation to the case of copolymers,
and to analyze the impact of disorder on mutual interactions between two star copolymers.

The questions of the influence of correlated disorder in the environment on the
scaling behavior of star copolymers and the resulting consequences remain so far unresolved and are
 the subject of the present study. We will also analyze the spectrum  of scaling exponents
  in particular for the  above mentioned process of trapping  diffusive particles
  in the vicinity of absorbing polymers in disordered environments.

The paper is organized as follows: in the next section we will give a field-theoretical
 representation of
the model Lagrangean. The field-theoretical renormalization group method, which
we use to find the qualitative characteristics of scaling behavior, is shortly described
in Section 3.
In Section 4 we discuss the results obtained. We finish by giving conclusions and an
outlook.

\section{The model}

Let us consider a polymer star with $f$ arms of different species in a
solvent.
We are working within the Edwards
continuous chain model \cite{Edwards65,Edwards66}, representing
each chain by a path $ r_i(s)$, parameterized by
$0\leq s\leq S_i$, $i=1,2,\ldots,f$.
 The central branching point of the star
is fixed at $ r_1(0)$.
The partition  function of the system is then defined by the
path integral
\cite{Schafer-1992}:
\begin{eqnarray}\nonumber
{\cal Z}_f(S_i)&=& \int D [  r_1,\ldots, r_f ] \\
\label{z} &&\times \exp \left[ -{\cal H}_f
\right]
\prod_{i=2}^f\delta^d( r_i(0)-r_1(0)).
\end{eqnarray}
Here, a multiple path integral is performed for the paths $r_1,\ldots,r_f$,
 the product of $\delta$-functions reflects the star-like configuration of $f$ chains,
 each starting at the point $r_1(0)$,
${\cal  H}_f$ is the Hamiltonian, describing the system of
$f$ disconnected polymer chains:
\begin{eqnarray} \label{eff}
{\cal H}_f&=&\frac{1}{2}\sum_{i=1}^f\int_0^{S_i}{\rm d}\,s
\left(\frac{{\rm d}\, r_{i}(s)}{{\rm d} s}\right)^2\nonumber\\
&&+\frac{1}{6}\sum_{i\leq j=1}^f  u_{ij}^0
\int {\rm d}  r \rho_i({\bf r} ) \rho_j({\bf r} ),
\end{eqnarray}
where $\rho_i(r)=\int_0^{S_i} {\rm d}
s\, \delta^d(r-r_{i}(s))$
and $u_{ij}^0$ is a symmetric matrix of bare excluded-volume interactions between
chains $i$ and $j$. 

The continuous chain model (\ref{z}) can be mapped onto a corresponding field theory
by a Laplace transform in the Gaussian surface
$S_i$ to the conjugated chemical potential variable (mass)
$\hat\mu_i$ \cite{Cloizeaux,Schafer91}:
\begin{equation} \label{laplace}
 \widehat{\cal Z}_{f}(\hat\mu_i)=\int\prod_{b}{\rm d}S_j
\exp[-\hat\mu_j S_j]{\cal Z}_{f}(S_i).
\end{equation}

One may then show that the Hamiltonian ${\cal H}$ is related to an
$m$-component field theory with a Lagrangean  ${\cal L}$
in the limit $m\to 0$:
\begin{eqnarray}
{ \cal L}\{\varphi_j,\mu_j\}&=&\frac{1}{2}
\sum_{i=1}^f\int{\rm d}^dx
 \left(\mu_i^2
|\vec{\varphi}_i(x)|^2+|\nabla\vec{\varphi}_i(x)|^2
\right)\nonumber\\
&+&\frac{1}{4!}\sum_{i\leq j=1}^f {u_{ij}^0} \int {\rm d}^d
x\, \varphi^{2}_i(x)\varphi^{2}_j(x),\label{holstar}
\end{eqnarray}
where $\varphi_i^m=\{\varphi_i^1,\ldots,\varphi_i^m\}$ and $\mu_i$
are bare critical masses.
On the base of the Lagrangean (\ref{holstar}) the one-particle irreducible vertex
functions $\Gamma^{(L)}$
of the theory can be obtained:
\begin{equation}
\delta(\sum_{q_i})\Gamma_{i_1,\ldots,i_L}^{{L}}(q_i)=\int{\rm e} ^{iq_ir_i} {\rm d}
 r_1 \cdots {\rm d} r_{L}
\langle
\varphi_{i_1}(r_1)\ldots \varphi_{i_L}(r_L) \rangle^{{\cal L}}_{1PI},
\end{equation}
where only those contributions, that have non-vanishing tensor factors in the limit
 $m\to 0$ are kept.

The Laplace-transformed partition function $ \widehat{\cal Z}_f(\hat\mu_i)$
has a vertex part, which is defined by the insertion of
the composite operator $\prod_i\varphi_i$:
\begin{eqnarray}
&&\delta(p+\sum_{j}q_j)\Gamma^{*f}(p,q_1,\ldots,q_f)=\label{com}\\
&&\int{\rm e} ^{i(pr_0+q_jr_j)}
{\rm d} r_0{\rm d} r_1\ldots {\rm d} r_{f}
\langle
\varphi_1(r_0)\cdots\varphi_f(r_0)
 \varphi_1(r_1)\cdots\varphi_f(r_f)
\rangle^{{\cal L}}_{1PI}.\nonumber
\end{eqnarray}

Let us note that we are interested in the case of a copolymer star, having
$f_1$ chains of one species and $f_2$ of another, so that $f_1+f_2=f$.
To keep notations simple we will consider in the following discussion
only two fields  $\varphi_1$ and $\varphi_2$, corresponding to two different
``species". Thus,
 in (\ref{holstar}) we have
interactions $u_{11}, u_{22}$ between the fields of the same ``species"
and $u_{12}$ between different fields.
The composite operator in  (\ref{com}) has the form of a product
$(\varphi_1)^{f_1}(\varphi_2)^{f_2}$.

We introduce disorder into the model (\ref{holstar}), by redefining
$\hat\mu_i^2 \to \hat\mu_i^2+\delta\hat\mu_i(x)$,  where the
 local fluctuations
$\delta\mu_i(x)$ obey:
$$
 \langle\langle\delta\mu_i(x)\rangle\rangle=0,
$$ $$
\langle\langle\delta\mu_i(x)\delta\mu_j(y)\rangle\rangle
=g_{ij}(|x-y|).
$$
Here, $\langle\langle\cdots\rangle\rangle$ denotes
the average over spatially homogeneous and isotropic quenched
disorder. The form of the pair correlation function $g(x)$ is
chosen to decay with distance according to the power law
(\ref{parcor}).

In order to average the free energy over different configurations
of the quenched disorder we apply the replica method to construct
an effective Lagrangean:
\begin{eqnarray}
&&{\cal L}_{eff}=\int {\rm d}x  \frac{1}{2}\sum_{i=1}^2\sum_{\alpha=1}^n\left[
(\vec{\nabla} \vec{\varphi}_i^{\alpha})^2 + \mu_i^2(\vec
{\varphi}_i^{\alpha})^2\right] \nonumber\\
&&+\sum_{i\leq j=1}^2 \sum_{\alpha=1}^n \frac{u_{ij}^0}{4!}
(\vec{\varphi}_i^{\alpha})^2(\vec{\varphi}_j^{\alpha})^2
\label{lag}\\&&
-\int {\rm d}x\,{\rm d}y \sum_{i\leq j=1}^2 \sum_{\alpha,\beta=1}^n g_{ij}(|x-y|)
(\vec{\varphi}_i^{\alpha})^2(\vec{\varphi}_j^{\beta})^2.\nonumber
\end{eqnarray}
Here, the coupling of the replicas is given by  the correlation
function  $g(x)$ of Eq.~(\ref{parcor}), Greek indices denote
replicas and the replica limit $n\to 0$ is implied.

For small $k$, the Fourier-transform $\tilde g_{ij}(k)$ of $g_{ij}(x)$
(\ref{parcor}) reads:
\begin{equation}
\label{fur}
\tilde g_{ij}(k)\sim {v}_{ij}^0+{w}_{ij}^0|k|^{a-d}.
\end{equation}
Thus, rewriting Eq.~(\ref{lag}) in momentum space, one obtains an effective Lagrangean
with 9 bare couplings: $u_{11}^0$, $u_{22}^0$, $u_{12}^0$, $v_{11}^0$, $v_{22}^0$,
$v_{12}^0$, $w_{11}^0$, $w_{22}^0$,
$w_{12}^0$.  As it was pointed out in Ref. \cite{Kim83}, once the
limit $m,n\to 0$ has been taken, the $u_{ij}^0$ and ${v_{ij}^0}$ terms acquire 
the same symmetry, and an effective Lagrangean with couplings
($u_{ij}^0-{v}_{ij}^0\equiv u_{ij}^0$) of $O(mn=0)$ symmetry  results.  This
leads to the conclusion that  weak quenched uncorrelated disorder
i.e. the case $a=d$ is irrelevant for polymers. Taking this into account, we end up
with only 6 couplings in an effective Lagrangean:
$u_{11}^0$, $u_{22}^0$, $u_{12}^0$, $w_{11}^0$, $w_{22}^0$, $w_{12}^0$.
For $a<d$, the momentum-dependent coupling ${w}_{ij}^0k^{a-d}$ has
to be taken into account.  Note that $\tilde g_{ij}(k)$ must be
positively definite being the Fourier image of the correlation
function. Thus, we have $w_{ij}^0>0$ for small $k$.
Note, that the couplings $u_{ij}^0$
should be positive, otherwise the pure system would undergo a
1st order transition.

The resulting Lagrangean in momentum space then reads:
\begin{eqnarray} \label{h}
&&{\cal L}_{\rm eff}= \frac{1}{2}
\sum_{\alpha=1}^n \sum_{i=1}^2 \sum_{k}\left[
k^2 + \mu_i^2\right]({\varphi}_i^{\alpha}(k))^2\nonumber\\
&&
+\sum_{i\leq j=1}^2 \mathop{\sum_{k_1k_2}}\limits_{k_3k_4} \left(\frac{u_{ij}^0}{4!}
\sum_{\alpha=1}^n
 \delta\,(k_1{+}k_2{+}k_3{+}k_4) \,
  {\vec{\varphi}
 _i^{\alpha}(k_1)\,\vec{\varphi}_{i}^{\alpha}(k_2)\,\vec{\varphi}_{j}^{\alpha}(k_3)\,
 \vec{\varphi}_{j}^{\alpha}
 (k_4)} \right. \label{model}
 \\
 &&
 - \left.\frac{w_{ij}^0}{4!}\sum_{\alpha, \beta=1}^n |k_1{+}k_2| ^{a - d }\delta\,
  ( k_1{+}k_2{-}
 k_3{ -}k_4)\,
\vec{\varphi}_{i}^{\alpha}(k_1)\,\vec{\varphi}_{i}^{\alpha}(k_2)\,\vec{\varphi}_{j}^{\beta}
(k_3)\, \vec{\varphi}_{j}^{\beta}(k_4)\right).\nonumber
\end{eqnarray}

In the next section, we apply the field-theoretical renormalization group approach
in order to extract the scaling behavior of the model (\ref{model}).

\section{Renormalization group approach}

We apply the renormalization group
(RG) method \cite{rg} in the massive
 scheme  renormalizing  the one-particle irreducible
vertex functions, in particular $\Gamma^{(2)},\Gamma^{(4)}$  and $\Gamma^{{2,1}}$, as well as
the vertex function $\Gamma^{*(f_1,f_2)}$, with a single
$(\varphi_1)^{f_1}(\varphi_2)^{f_2}$ insertion.
Note that  the polymer limit of a zero component field leads to
an essential simplification: each field $\varphi_i$, mass $\mu_i$ and coupling $u_{ii}^0$
renormalizes as if the other fields were absent.
The renormalized couplings $u_{ij},w_{ij}$ are given by:
\begin{eqnarray}
u_{ii}^0=\mu^{\varepsilon}Z_{\varphi_i}^{-2}Z_{ii}u_{ii},\,\,\, \,\,\,\,i=1,2\\
w_{ii}^0=\mu^{\delta}Z_{\varphi_i}^{-2}Z_{ii}w_{ii},\,\,\,\,\,\,\, i=1,2\\
u_{12}^0=\mu^{\varepsilon}Z_{\varphi_1}^{-1}Z_{\varphi_2}^{-1}Z_{12}u_{12},\\
w_{12}^0=\mu^{\delta}Z_{\varphi_1}^{-1}Z_{\varphi_2}^{-1}Z_{12}w_{12}.
\end{eqnarray}
Here, $\mu$ is a scale parameter, equal to the renormalized mass, and parameters
 $\varepsilon=4-d$, $\delta=4-a$. The renormalization factors
 $Z$ have the form of power series, the coefficients of which are calculated
perturbatively order by order.


The star vertex function $\Gamma^{*(f_1,f_2)}$ is renormalized by a factor $Z_{*f_1,f_2}$:
\begin{equation}\label{3.8}
Z_{\varphi_1}^{f_1/2}Z_{\varphi_2}^{f_2/2}Z_{* f_1,f_2}
\Gamma^{(* f_1 f_2)}
 = \mu^{(f_1+f_2)(\varepsilon/2-1)+4-\varepsilon} .
\end{equation}

The variation of the coupling constants under renormalization
defines a flow in parametric space, governed by corresponding
$\beta$-functions:
\begin{eqnarray}
\beta_{u_{ij}}(u_{ij},w_{ij})=\mu \frac{\rm d}{{\rm d}\mu}u_{ij},\,\,\,
\beta_{w_{ij}}(u_{ij},w_{ij})=\mu \frac{\rm d}{{\rm d}\mu}w_{ij},\,\,i,j=1,2.
\end{eqnarray}
The fixed points (FPs) of the RG transformation are
given by the  solution of the system of equations:
\begin{equation}
\beta_{u_{ij}}(u_{ij}^*,w_{ij}^*)=0,\,\,\,
\beta_{w_{ij}}(u_{ij}^*,w_{ij}^*)=0,\,\,i,j=1,2. \label{FP}
\end{equation}
The stable FP, corresponding to the
critical point of the system, is defined as the fixed point where
the stability matrix
possesses eigenvalues $\{\lambda_i\}$ with positive real parts.

The flow of the renormalizing factors $Z$ in turn gives rise to RG functions
$\eta_{\varphi_i}$ and $\eta_{*f_1f_2}$ as follows:
\begin{eqnarray}
\mu \frac{\rm d}{{\rm d}\mu} \ln Z_{\varphi_i} &=&
\eta_{\varphi_i}(u_{ij},w_{ij}),                        \label{3.13}\\
\mu \frac{\rm d}{{\rm d}\mu} \ln Z_{*f_1f_2} &=&
\eta_{*f_1f_2}(u_{ij},w_{ij}).                     \label{3.14}
\end{eqnarray}
At the FP of the renormalization group transformation, the function $\eta_{\varphi_i}$
 describes the pair correlation function
critical exponent,
while the functions $\eta_{*f_1f_2}$
define the set of exponents for copolymer stars:
\begin{eqnarray}
&& \eta = \eta_{\varphi_i}(u_{ij}^*,w_{ij}^*)\\
&& \eta_{f_1f_2} =\eta_{*f_1f_2}(u_{ij}^*,w_{ij}^*).
\end{eqnarray}
In the next section, we will present
expressions for the $\beta$
and $\eta$ functions,  together with a
study of the RG flow and the fixed points of the theory.

\section{The results}

\subsection{Fixed points and scaling exponents}
According to the renormalisation group prescriptions, we obtain the RG
functions of the model
(\ref{model})
within a massive scheme up to the one-loop approximation:
\begin{eqnarray}
&&\beta_{u_{ii}}=-\varepsilon\left[u_{ii}-\frac{4}{3}u_{ii}^2I_1\right]-\delta
2u_{ii}w_{ii}\left[I_2+\frac{1}{3}I_4\right]+
(2\delta-\varepsilon)w_{ii}^2I_3,\\
&&\beta_{w_{ii}}=-\delta\left[w_{ii}+\frac{2}{3}w_{ii}^2I_2+\frac{2}{3}w_{ii}^2I_4\right]+
\varepsilon\frac{2}{3}w_{ii}u_{ii}I_1, \,\,\,\,\,\,\,\,\,i=1,2;\\
&&\beta_{u_{12}}=-\varepsilon\left[u_{12}-\frac{2}{3}u_{12}^2I_1-\frac{1}{3}u_{12}(u_{11}+u_{22})I_1^2\right]\nonumber\\
&& -
\delta \left[u_{12}w_{12}I_2+\frac{1}{2}u_{12}(w_{11}+w_{22})I_2+\frac{1}{2}u_{12}(w_{11}+w_{22})I_4)\right]
\nonumber\\
&&+
(2\delta-\varepsilon)\left[\frac{1}{3}w_{12}^2I_3+
\frac{1}{6}w_{12}(w_{11}+w_{2})\right],\\
&&\beta_{w_{12}}=-\delta\left[w_{12}+\frac{1}{3}w_{12}^2I_2+\frac{1}{3}w_{ii}^2I_4\right]\nonumber\\
&&+
\varepsilon\left[\frac{1}{3}w_{12}u_{12}I_1+\frac{1}{6}w_{12}(u_{11}+u_{22})I_1+\frac{1}{6}w_{12}(w_{11}+
w_{12})I_2)\right].
\end{eqnarray}
Note, that expressions for $\beta_{u_{ii}},\beta_{w_{ii}}$
restore the corresponding RG functions
for a single polymer chain in long-range correlated disorder \cite{Blavatska01,Blavatska01a}.
Here, $I_i$ are the loop-integrals:
\begin{eqnarray}
&&I_1=\int\frac{{\rm d}{\vec q}}{(q^2+1)^2},\nonumber\\
&&I_2=\int\frac{{\rm d}{\vec q}\,q^{a-d}}{(q^2+1)^2},\nonumber\\
&&I_3=\int\frac{{\rm d}{\vec q}\,q^{2(a-d)}}{(q^2+1)^2},\nonumber\\
&&I_4=\frac{\partial }{\partial k^2}\left [ \int\frac{
{\rm d}\vec{q}\, q^{a-d}}{[q+k]^2 + 1) }\right ]_{k^2=0}.
\end{eqnarray}
 We make the couplings dimensionless by
redefining $u_{ij}=u_{ij}\mu^{d-4}$ and
$w_{ij}=w_{ij}\mu^{a-4}$, therefore, the loop integrals do not explicitly
contain the mass. Besides, we absorb geometrical factors $S_d$, resulting
 from angular integration into the couplings.

Additionally, we need the RG function $\eta_{*f_1f_2}(u_{ij},w_{ij})$, which we find in
the form:
\begin{eqnarray}
&&\eta_{*f_1f_2}=-\varepsilon\left(u_{11} \frac{f_1(f_1-1)}{6}I_1+u_{22}
\frac{f_2(f_2-1)}{6}I_1
+u_{12}\frac{f_1f_2}{3}I_1\right)+\nonumber\\
&&+\delta\left(w_{11}\frac{f_1(f_1-1)}{6}I_2+
w_{22}\frac{f_2(f_2-1)}{6}I_2+w_{12}\frac{f_1f_2}{3}I_2\right).
\end{eqnarray}

The perturbative expansions for RG functions may be analyzed
by two complementary approaches: either by exploiting a double expansion in
the parameters $\varepsilon=4-d,
\delta=4-a$ \cite{Blavatska01,Blavatska01a,Weinrib83},  or by fixing
the values of the parameters $d,a$
\cite{Blavatska01}.
 Let us note, that  within the one-loop approximation the latter method
 can not give reliable results \cite{Blavatska01},
and we exploit the double expansion in
 $\varepsilon=4-d$, $\delta=4-a$ for a qualitative analysis.
The resulting expressions for $\beta$- and $\eta$-functions read: \begin{eqnarray}
&&\beta_{u_{ii}}=-\varepsilon
u_{ii}+\frac{4}{3}u_{ii}^2-2u_{ii}w_{ii}+\frac{2}{3}w_{ii}^2 \label{beta1}\\
&&\beta_{w_{ii}}=-\delta w_{ii}-\frac{2}{3}w_{ii}^2+\frac{2}{3}u_{ii}w_{ii}, \,\,\,\,\,i=1,2; \\
&&\beta_{u_{12}}=-\varepsilon
u_{12}+\frac{2}{3}u_{12}^2+\frac{1}{3}u_{12}(u_{11}+u_{22})-
\frac{1}{2}u_{12}(w_{11}+w_{22})\nonumber\\
&&-
u_{12}w_{12}+\frac{1}{3}w_{12}^2+\frac{1}{6}w_{12}(w_{11}+w_{22}), \label{beta2}\\
&&\beta_{w_{12}}=-\delta
w_{12}-\frac{1}{3}w_{12}^2+\frac{1}{3}u_{12}w_{12}\nonumber\\
&&+\frac{1}{6}w_{12}(u_{11}+
u_{22})-\frac{1}{6}w_{12}(w_{11}+w_{22}),\label{beta3}\\
&&\eta_{*f_1f_2}=-\frac{f_1(f_1-1)}{6}(u_{11}-w_{11})-\frac{f_2(f_2-1)}{6}(u_{22}
-w_{22})\nonumber\\
&&-\frac{f_1(f_1-1)}{3} u_{12}+
\frac{f_2(f_2-1)}{3}w_{12}.\label{etaexp}
\end{eqnarray}
Substituting Eqs. (\ref{beta1})-(\ref{beta3}) into (\ref{FP}), we find a number of
fixed points,
corresponding to  different scenarios of
the scaling  behavior of the model.

{\it Pure solution}
First, let us consider the case, when disorder is absent ($w_{11}=w_{22}=w_{12}=0$) and we
recover
the problem of the so-called ternary solution of two polymer species in a good solvent
 \cite{Schafer91}.
 Solving the equations $\beta_{u_{ij}}=0$, $i,j=1,2$, we
find eight fixed points in correspondence with Refs. \cite{Ferber97,Ferber97e,Ferber99}.
The trivial FPs: $G_0 (u^*_{11}=u^*_{22}=u^*_{12}=0)$,
 $U_0  (u^*_{11}\neq 0,u^*_{22}=u^*_{12}=0)$,
$U^{'}_0 (u^*_{22}\neq 0,u^*_{11}=u^*_{12}=0)$ and
$S_0 (u^*_{11}=u^*_{22}\neq 0,u^*_{12}=0)$ describe sets of two
mutually non-interacting polymer species. More
interesting are the FPs denoted as $G$, $U$, $U^{'}$, $S$,  describing two
mutually interacting species, their coordinates are given in the upper part of
Table \ref{tab1}.
 Corresponding values of the  exponents $\eta_{f_1f_2}$ read:
\begin{eqnarray}
\eta_{f_1f_2}^{G}&=&\frac{-(f_1f_2)\varepsilon}{2},\nonumber\\
\eta_{f_1f_2}^{U}&=&\eta_{f_2,f_1}^{U^{'}}=\frac{-f_1(f_1+3f_2-1)\varepsilon}{8},\nonumber\\
\eta_{f_1f_2}^{S}&=&\frac{-(f_1+f_2)(f_1+f_2-1)\varepsilon}{8}.
\label{etapure}
\end{eqnarray}
Note, that $\eta_{f_1f_2}^{S}$ just recovers the exponent of a homogeneous polymer star
with $f=f_1+f_2$ arms.
The values of these exponents are known up to 4th order of
the $\varepsilon$-expansion \cite{Schafer-1992,Schulte04}
and in the fixed $d$ approach \cite{Ferber97}.
\begin{table}[b!]
\caption{ \label{tab1} Non-trivial fixed points of the model (\ref{h}).}
\begin{tabular}{ l l l  l  l  l l }
\hline
\hline
 & $u_{11}^*$ &$u_{22}^*$& $u_{12}^*$ & $w_{11}^*$& $w_{22}^*$ & $w_{12}^*$\\
\hline
$G$ & 0 & 0 & $\frac{3\varepsilon}{2}$ &0 &0 &0 \\
$U$ & $\frac{3\varepsilon}{4}$ & 0 & $\frac{9\varepsilon}{8}$ & 0 &0 & 0\\
$U^{'}$ & 0 & $\frac{3\varepsilon}{4}$ & $\frac{9\varepsilon}{8}$ & 0 &0 & 0\\
$S$ & $\frac{3\varepsilon}{4}$ & $\frac{3\varepsilon}{4}$ & $\frac{3\varepsilon}{4}$& 0 &0 & 0\\
 $G_{L}$ & 0 & 0 &$\frac{3\delta^2}{(\varepsilon-\delta)}$ &0 & 0  &  $\frac{3\delta(\varepsilon-2\delta)}{(\delta-\varepsilon)}$\\
 $U_{L}$ & $\frac{3\delta^2}{2(\varepsilon-\delta)}$ & 0 & $\frac{9\delta^2}{4(\varepsilon-\delta)}$
 & $\frac{3\delta(\varepsilon-2\delta)}{2(\delta-\varepsilon)}$ & 0 & $\frac{9\delta(2\delta-\varepsilon)}{4(\varepsilon-
\delta)}$\\
 $U^{'}_{L}$ & 0 &$\frac{3\delta^2}{2(\varepsilon-\delta)}$ & $\frac{9\delta^2}{4(\varepsilon-\delta)}$
& 0 & $\frac{3\delta(\varepsilon-2\delta)}{2(\delta-\varepsilon)}$  & $\frac{9\delta(2\delta-\varepsilon)}{4(\varepsilon-
\delta)}$\\
$S_{L}$
 & $\frac{3\delta^2}{2(\varepsilon-\delta)}$
 & $\frac{3\delta^2}{2(\varepsilon-\delta)}$
 & $\frac{3\delta^2}{2(\varepsilon-\delta)}$
 & $\frac{3\delta(\varepsilon-2\delta)}{2(\delta-\varepsilon)}$
 & $ \frac{3\delta(\varepsilon-2\delta)}{2(\delta-\varepsilon)}$
 &  $\frac{3\delta(\varepsilon-2\delta)}{2(\delta-\varepsilon)}$ \\
\hline
\hline
\end{tabular}
\end{table}

{\it Solution in the presence of long-range correlated disorder}. Next, let us turn on the disorder.
Apart from the eight FPs listed above, now we have a whole set of new FPs describing two polymer species
in the case, when one or both of the species feel the presence of long-range correlated disorder.
Indeed, to find these FPs one has to solve the system of 6 second-order equations (\ref{FP})
with the $\beta$-functions given by $(\ref{beta1})-(\ref{beta3})$. In principle, this may
lead to $2^6$ solutions \cite{FP}.
 In what follows we consider only four nontrivial points, corresponding to
copolymer stars of mutually interacting species, both feeling the presence of disorder,
which are of foremost interest (see Table \ref{tab1}). These FPs describe particular
 situations of two mutually interacting sets of RWs
($G_L$), SAWs ($S_L$)  and two interacting sets of RWs and SAWs ($U_L$, $U^{'}_{L}$).
Note that due to the special form of the $\beta$-functions the fixed points with
$u_{ii}^*= 0$, $w_{ii}^*\neq 0$ do not exist, i.e. one cannot
describe simple random walks in the media with
long-range-correlated disorder.

We are interested in the points, which are stable in all coordinate directions.
After analyzing the stability and physical accessibility of all the points,
we come to the conclusion, that only the FPs $S$ and $S_{L}$  are stable in all
 directions and their stabilities are
determined by the conditions:
\begin{itemize}
\item fixed point $S$ is stable for $\varepsilon>2\delta$,
\item fixed point $S_{L}$ is stable for $\delta<\varepsilon<2\delta$.
\end{itemize}
Although the remaining FPs ($G_{L}$, $U_{L}$ and $U^{'}_{L}$ from the Table 1) are
unstable, they can be
reached for $\delta<\varepsilon<2\delta$ under specific
initial conditions. In particular, $G_L$ is reachable
from the initial condition $u_{11}=u_{22}=w_{11}=w_{22}=0$, $U_{L}$ is reachable
for $u_{22}=w_{22}=0$
and $U^{'}_{L}$ for $u_{11}=w_{11}=0$. Substitution of these FPs values into
the expansion (\ref{etaexp})
results in the following estimates for
$\eta_{f_1f_2}$:
\begin{eqnarray}
\eta^{G_{L}}_{f_1f_2}&=&-(f_1f_2)\delta,\nonumber\\
\eta^{U_{L}}_{f_1f_2}&=&\eta^{U^{'}_{L}}_{f_2,f_1}= \frac{-f_1(f_1+3f_2-1)\delta}{4},\nonumber\\
\eta^{S_{L}}_{f_1f_2}&=& \frac{-(f_1+f_2)(f_1+f_2-1)\delta}{4}.
\label{etalr}
\end{eqnarray}
Here, $\eta^{S_{L}}_{f_1f_2}$ gives the exponent for the homogeneous star
with $f_1+f_2$ arms in solution in long-range-correlated disorder,
 $\eta^{G_{L}}_{f_1f_2}$ and $\eta^{U_{L}}_{f_1f_2}$ describe $f_2$ random walks,
interacting with $f_1$ RWs and with $f_1$ SAWs respectively,
in long-range-correlated disorder. All this leads to a variety of
 new scaling behavior for copolymer stars in a disordered medium.

\subsection{Diffusion-limited reaction rates}

Let us consider the $f_1$-arm star polymer with arms of linear size $R_s$ and 
absorbing sites all along these arms. 
 At the center of the star a particular absorbing trap is placed.
Free particles $A$ which  diffuse in solution are trapped or react at these sites.
We are interested in the reaction rate $k_{f_2}$ of simultaneously
trapping $f_2$ randomly walking particles $A$. This rate is proportional to the averaged moments
of the concentration $\rho$ of the particles near this trap and scales as
\cite{Cates87,Ferber97e,Ferber97,Ferber99}:
\begin{equation}
k_{f_2}\sim \langle \rho^{f_2}\rangle \sim R_s^{-\lambda_{f_1f_2}}.
\label{rho}
\end{equation}
This process is an example of a so-called diffusion-limited reaction \cite{Rice,Burlatsky}, with the
rate depending on the sum of the
diffusion coefficients of the reactants \cite{Berg85}. As far as the presence of
disorder lowers the
diffusion coefficients \cite{Minton, Zimmerman93}, it is predicted to
lower rates of association in
 diffusion-limited circumstances. It is interesting to
check this prediction analytically, analyzing the behavior of star copolymers in
long-range correlated disorder.
In terms of the path integral solution of the diffusion equation, one finds that
to calculate the rate of a reaction at the absorber that involves $f_2$
particles simultaneously one needs to consider $f_2$ RWs that end at this point.
The moments of concentration in Eq. (\ref{rho}) are thus defined by a
partition function of a star comprising $f_2$ RWs
\cite{Duplantier-1989I,Schafer-1992}.
Finally, introducing the mutual avoidance conditions between the absorbing star
and a ``star'' of diffusive particles one
ends up with the problem of calculating the partition function of a copolymer
star with two species $f_1$, $f_2$.
By means of the short-chain expansion \cite{Ferber97c} the set of exponents $\eta_{f_1f_2}$ in (\ref{comstar}) can be related to the
exponents $\lambda_{f_1f_2}$ in (\ref{rho}) \cite{Ferber97e,Ferber97,Ferber99,Ferber01}:
\begin{eqnarray} \label{etalambda}
\lambda^{RW}_{f_1f_2} =-\eta^{G}_{f_1f_2}, \nonumber\\
\lambda^{SAW}_{f_1f_2} =
-\eta^{U}_{f_1f_2}+\eta^{U}_{f_1 0}.
\end{eqnarray}
Based on these relations,  the resulting values for the pure solution read \cite{Ferber01}:
\begin{eqnarray}
\lambda^{RW_{pure}}_{f_1f_2}=\frac{\varepsilon}{2}f_1f_2,\nonumber \\
\lambda^{SAW_{pure}}_{f_1f_2}=\frac{3 \varepsilon}{8}f_1f_2. \label{lambdapure}
\end{eqnarray}
Let us note, that the case $f_1=2$ corresponds to a trap
located on the chain polymer, whereas $f_1=1$ corresponds to a trap
attached at the polymer extremity.

Corresponding values for the exponents defining these processes in an environment
with long-range correlated disorder
can be obtained by substituting Eqs. (\ref{etalr}) into (\ref{etalambda}):
\begin{eqnarray}
\lambda^{RW_L}_{f_1f_2}=-\eta^{G_{L}}_{f_1f_2}=\delta f_1f_2,\nonumber\\
\lambda^{SAW_L}_{f_1f_2}=-\eta^{U_{L}}_{f_1f_2}+\eta^{U}_{f_1 0}=\frac{3 \delta}{4}f_1f_2.
\label{lambdalr}
\end{eqnarray}

Comparing relations (\ref{lambdapure}) and (\ref{lambdalr}) at
fixed values $\varepsilon=1$ $(d=3)$ and varying the parameter $\delta$, one notes that
the presence of correlated disorder results in an increase of
the exponents $\lambda$. Moreover, the stronger the correlation of
defects, the larger is
$\lambda$.
Recalling the definition (\ref{rho}), we immediately conclude that, as expected,
the  presence of long-range correlated disorder results in {\it lowering} the rates of diffusion-limited reactions.
The crucial point is that while long-range correlated disorder
apparently does not influence the RW itself (there is no new fixed point with $u_{ii}=0$, $w_{ii}\neq 0$),
the fact that the absorbing polymer changes its conformation and
fractal dimension in the LR background leads to a change of the diffusive
behavior of particles being absorbed (or catalyzed) on the polymer.

Let us analyze several particular cases:
\begin{itemize}
\item
 For a given $f_1$-star absorber i.e. a reactive site placed at one end of
an otherwise absorbing polymer
 increasing
the size $R_s$ by a factor of $l$ changes the reaction rate to
$k^{\prime}_{f_1f_2} \sim (lR_s)^{-\lambda_{f_1f_2}}$,
so that:
\begin{equation}
k^{\prime}_{f_1f_2}/k_{f_1f_2} \sim l^{-\lambda_{f_1f_2}}.
\end{equation}
Increasing the size of the polymer thus
leads to a reaction rate decrease by a factor of $l^{-\lambda_{f_1f_2}}$ .
Since $\lambda^{L}_{f_1f_2}$ is larger than $\lambda^{pure}_{f_1f_2}$, we conclude,
that the presence of long-range correlated defects
makes the reaction rate decreases more slowly as compared to the pure solution case.
\item
For a fixed number $f_2$ of
particles to be trapped simultaneously the effect of attaching $f^{'}_1$  additional arms
to an $f_1$-arm star absorber decreases the reaction rate:
\begin{equation}
k_{f_1+f^{'}_1\,f_2}/k_{f_1f_2}
\sim R_s^{-\left(\lambda_{f_1+f^{'}_1\,f_2} -\lambda_{f_1f_2}\right)},
\end{equation}
as far as $\lambda_{f_1+f^{'}_1\,f_2}>\lambda_{f_1f_2}$. This decrease is
suppressed to some extent   in the presence of long-range correlated defects.

\item For a given $f_1$-star absorber an increase of the number of  particles
 to be trapped simultaneously  results in a decrease of the reaction rate:
\begin{equation}
k_{f_1 f_2+f^{'}_2}/k_{f_1f_2} \sim R_s^{-\left(\lambda_{f_1 f_2+f^{'}_2} -\lambda_{f_1f_2}\right)},
\end{equation}
since $\lambda_{f_1 f_2+f^{'}_2}>\lambda_{f_1f_2}$. Again, presence of disorder makes the reaction rate decrease
more slowly as compared to pure case.
\end{itemize}

\subsection{Interaction between star copolymers}
\begin{figure}[b!]
 \begin{center}
\caption{\label{contact} (color online) Three non-trivial examples of copolymer stars where the interaction is governed by contact exponents
$\Theta^{S\,\,S}_{f_1f_2 \,\,g_1g_2}$ (a), $\Theta^{U\,\,U}_{f_1f_2 \,\,g_1g_2}$ (b) and $\Theta^{G\,\,G}_{f_1f_2 \,\,g_1g_2}$ (c).}
\includegraphics[width=4.4cm]{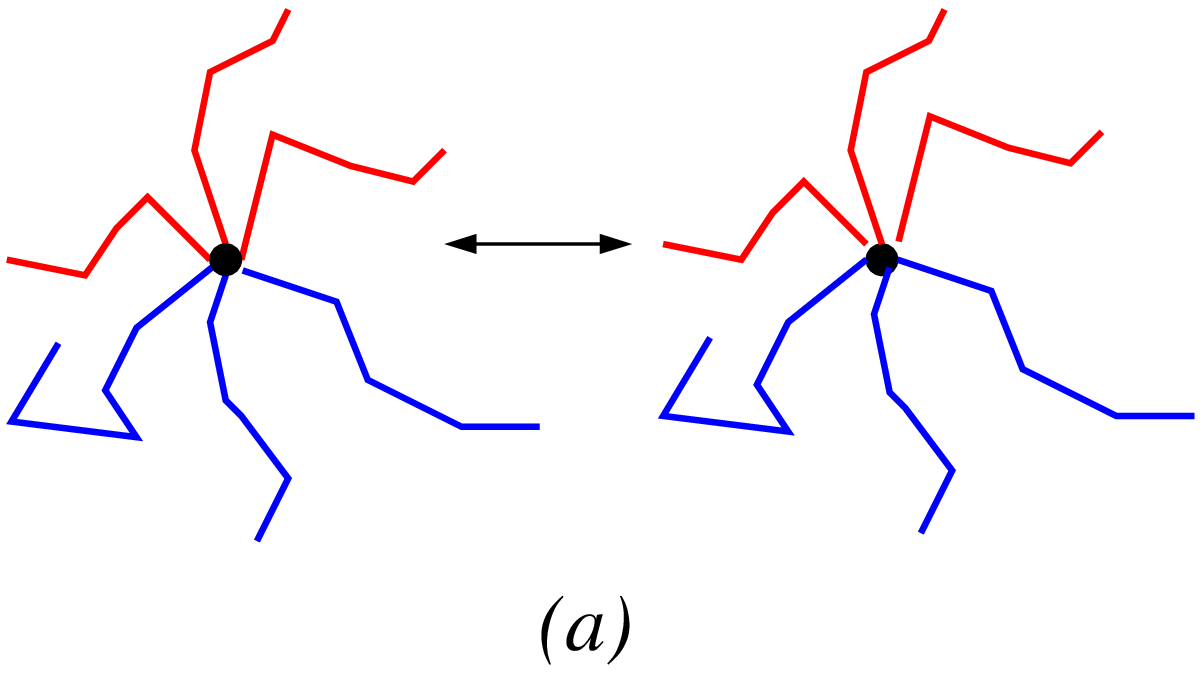}\hspace*{1cm}
\includegraphics[width=4.5cm]{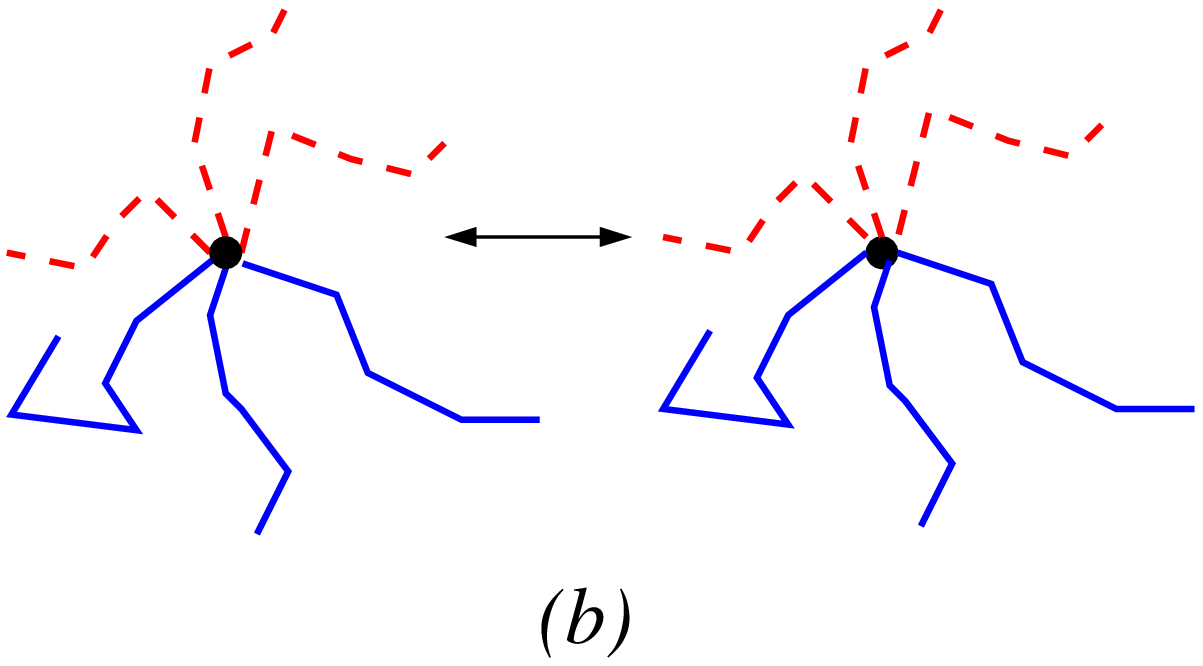}\hspace*{1cm}
\includegraphics[width=4.5cm]{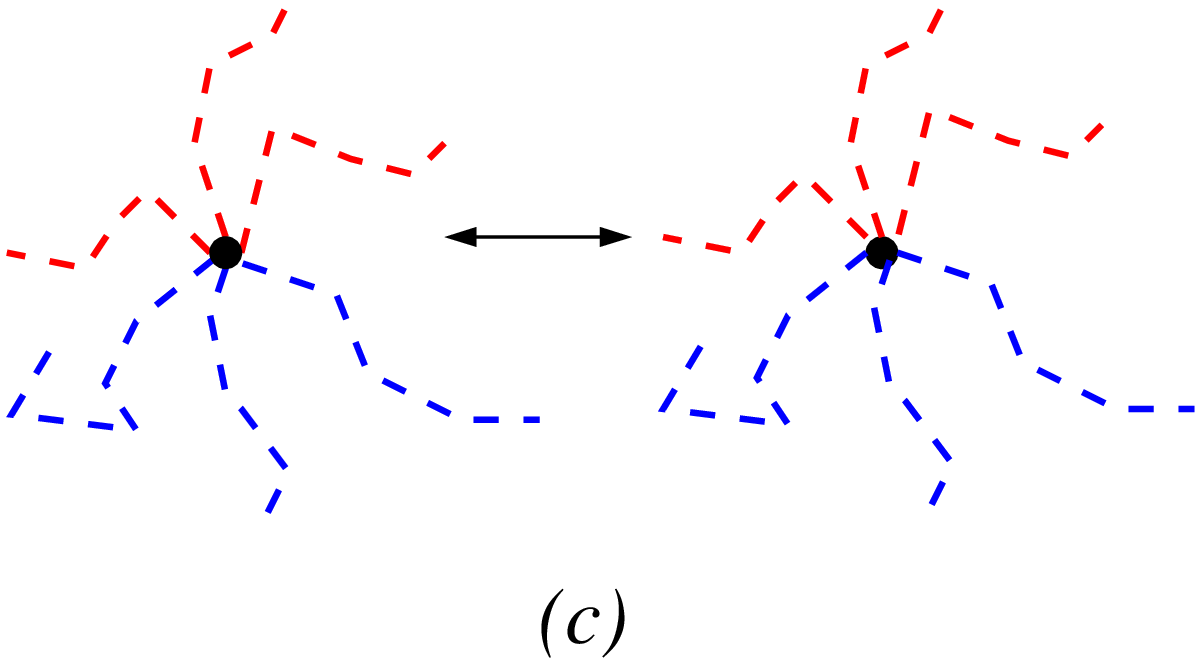}
\end{center}
\end{figure}
The effective interaction between two star copolymers at short distance $r$ between their cores can be
estimated following the scheme of Refs. \cite{Duplantier-1989I,Ferber97c,Ferber00}, based on short distance expansion. The partition sum
$Z_{f_1f_2\,\,g_1g_2}$(r) of the two stars with 
$f=f_1+f_2$ and $g=g_1+g_2$ arms of species $1$ and $2$ at small center-to-center distances $r$ factorizes into a
function $C_{f_1f_2\,\,g_1g_2}(r)$ and the
partition function $Z_{f_1+g_1\,\,f_2+g_2}$ of the star with $f_1 + g_1$ arms of species $1$ and $f_2+g_2$ arms of species $2$
which is formed when the cores of the two stars coincide:
\begin{equation}
Z_{f_1f_2\,\,g_1g_2}(r) \simeq C_{f_1f_2\,\,g_1g_2}(r)Z_{f_1+g_1\,f_2+g_2}.
\label{4}
\end{equation}
For the function $C_{f_1f_2\,\,g_1g_2}(r)$
it was shown \cite{Duplantier-1989I,Ferber97c} that power-law scaling for small $r$ holds in the form:
\begin{equation}
C_{f_1f_2\,\,g_1g_2}(r) \simeq r^{\Theta_{f_1f_2\,\,g_1g_2}}.
\end{equation}
To find the scaling relation for this power law, we take into account (\ref{comstar}) and change the length scale in an invariant way
by: $r\to \ell r$, $R\to\ell R$. Eq. (\ref{4}) then can be written as:
\begin{eqnarray}
\ell ^{\eta_{f_1f_2}-f_1\eta_{20}-f_2\eta_{02}}\ell^{\eta_{g_1g_2}-g_1\eta_{20}-
g_2\eta_{02}}Z_{f_1f_2\,\,g_1g_2}(r)=
\nonumber\\
\ell^{\Theta_{f_1f_2\,\,g_1g_2}}\ell^{\eta_{f_1+g_1\,f_2+g_2}-
(f_1+g_1)\eta_{20}-(f_2+g_2)\eta_{02}}Z_{f_1+g_1\,\,f_2+g_2}.
\end{eqnarray}
Collecting powers of $\ell$ provides the scaling relation for the contact exponent:
\begin{eqnarray}
&&\Theta_{f_1f_2\,\,g_1g_2} =
\eta_{f_1f_2}-f_1\eta_{20}-f_2\eta_{02} + \eta_{g_1g_2}-g_1\eta_{20}-g_2\eta_{02}-\\ \nonumber
&& (\eta_{f_1+g_1\,f_2+g_2}-
(f_1+g_1)\eta_{20}-(f_2+g_2)\eta_{02})=\\ \nonumber
&&\eta_{f_1f_2} + \eta_{g_1g_2}- \eta_{f_1+g_1\,f_2+g_2}.\label{teta}
\label{}
\end{eqnarray}
For two star copolymers at a distance $r$  between their centers
 the mean force $F_{f_1f_2\,\,g_1g_2}(r)$ acting
on the centers can be derived as the gradient of
the effective potential $V_{eff}(r) = -k_JT \log[Z_{f_1f_2\,\,g_1g_2}(r)/(Z_{f_1g_1}Z_{f_2g_2})]$.
 For the force at short distances $r$ this  results in \cite{Ferber01a}:
\begin{equation}
F_{f_1f_2\,\,g_1g_2}(r)=k_JT\frac{\Theta_{f_1f_2\,\,g_1g_2}}{r}.
\end{equation}

For two mutually interacting star copolymers we have the three following nontrivial situations.
First one may have two stars, each
consisting of two species (with numbers of arms $f_1$, $f_2$ and $g_1$, $g_2$ respectively) 
all behaving as mutually avoiding SAWs (see Fig. \ref{contact}a). This situation is equivalent to two SAW star polymers 
of the same species. 
A second  possible situation is the interaction between two star copolymers, the first
 containing $f_1$ SAWs and $f_2$ RWs,
another $g_1$ SAWs and $g_2$ RWs (Fig. \ref{contact}b). Thirdly, one may  have two stars, each
consisting of two species (with $f_1$, $f_2$ and $g_1$, $g_2$ arms respectively) all behaving like RWs but with mutual avoidance between the species (Fig. \ref{contact}c).
It is easy to check, that any other case can be represented in terms of these three nontrivial examples. E.g.,
putting $f_2=0$ in the case corresponding to Fig. \ref{contact}b, we obtain a homogeneous $f_1$-arm star polymer
interacting with a star copolymer, etc.

\begin{figure}[t!]
\caption{\label{th} Contact exponents of interaction between a copolymer star with $f_1$ SAWs and $f_2$ RWs and:
an $8$-armed star of RWs(a); an $8$-armed star of SAWs(b); a copolymer star with $4$ arms of SAWs and $4$ arms of RWs in $d=3$.
Boxes: pure case ($a=3$), circles: $a=2.7$, triangles: $a=2.2$}
 \begin{center}
\includegraphics[width=6.5cm]{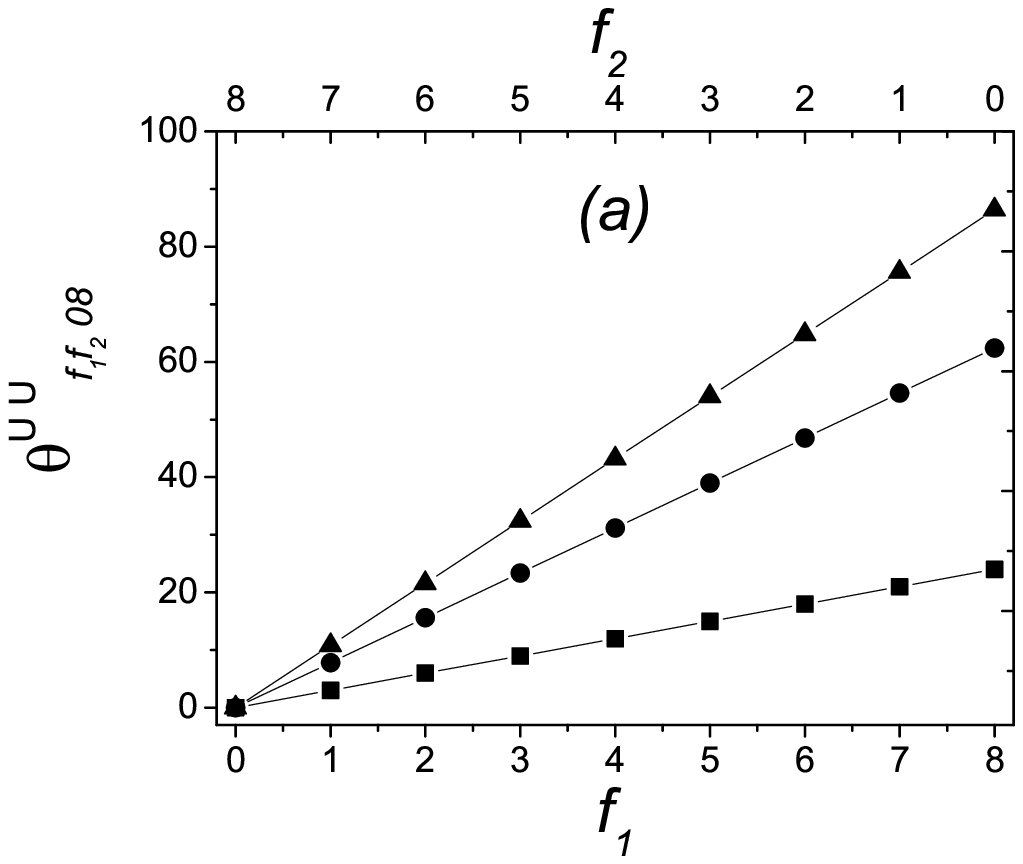}

\includegraphics[width=6.5cm]{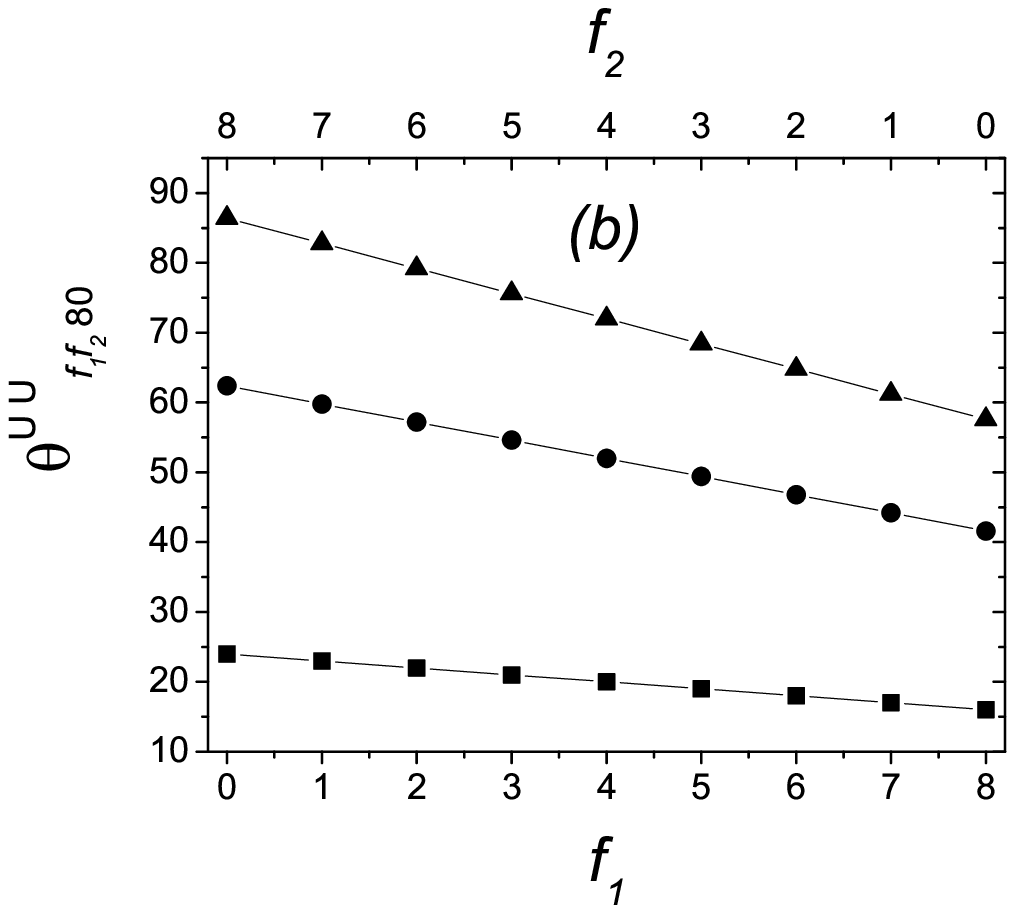}

\includegraphics[width=6.5cm]{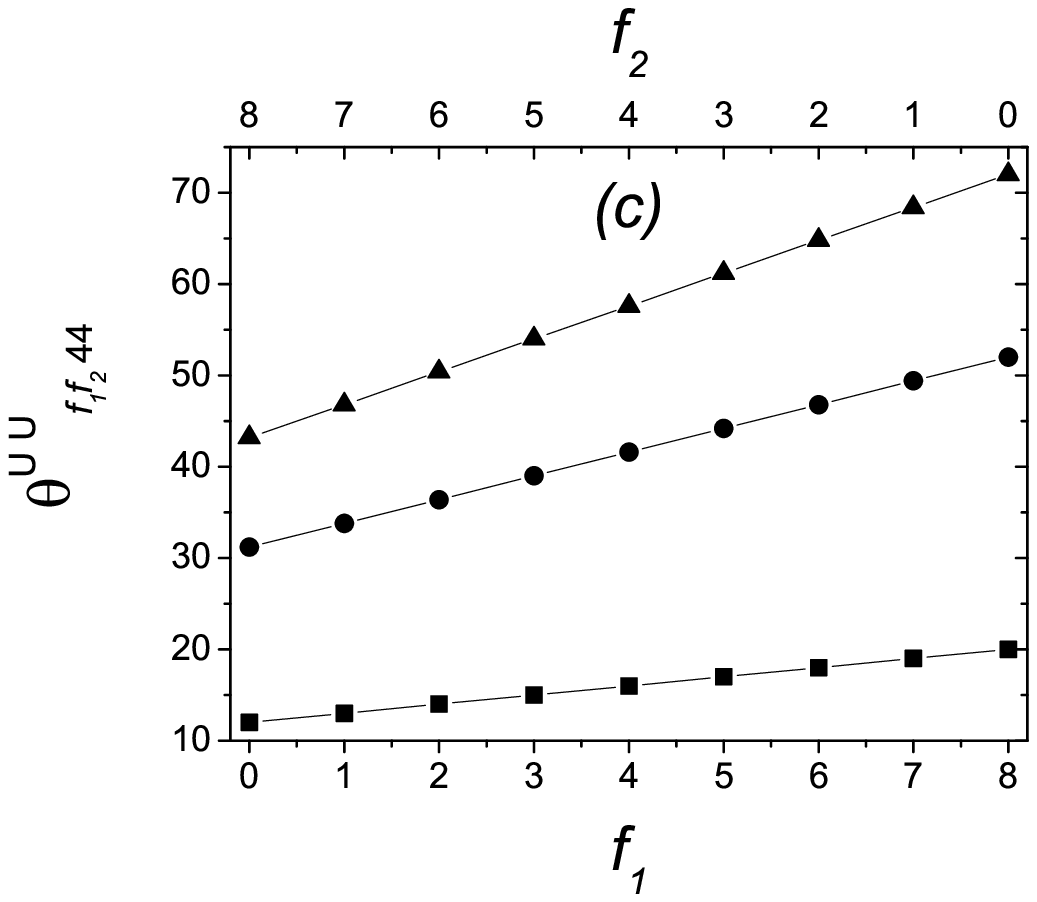}
\end{center}
 \end{figure}

Taking into account Eqs. (\ref{etapure}), (\ref{etalr}) we find the following contact exponents corresponding
to the three nontrivial situations described above.

{\it 1) Pure Solution}
\begin{eqnarray}
&&\Theta^{S\,\,S}_{f_1f_2 \,\,g_1g_2}=\eta^{S}_{f_1+f_2}+\eta^{S}_{g_1+g_2}-\eta^S_{f_1+f_2+g_1+g_2}=
\frac{\varepsilon}{4}(f_1+f_2)(g_1+g_2)\\
&&\Theta^{U\,\,U}_{f_1f_2\,\,g_1g_2}=\eta^{U}_{f_1f_2}+\eta^{U}_{g_1g_2}-
\eta^{U}_{f_1+g_1\,f_2+g_2}=\frac{\varepsilon}{8}(2f_1g_1+3f_1g_2+3g_1f_2)\\
&&\Theta^{G\,\,G}_{f_1f_2\,\,g_1g_2}=\eta^{G}_{f_1f_2}+\eta^{G}_{g_1g_2}-
\eta^{G}_{f_1+g_1\,f_2+g_2}=\frac{\varepsilon}{2}(f_1g_2+f_2g_1).
\end{eqnarray}
{\it 2) Presence of LR disorder}
\begin{eqnarray}
&&\Theta^{(S\,\,S)_L}_{f_1f_2 \,\,g_1g_2}=\eta^{S_L}_{f_1+f_2}+\eta^{S_L}_{g_1+g_2}-\eta^{S_L}_{f_1+f_2+g_1+g_2}=
\frac{\delta}{2}(f_1+f_2)(g_1+g_2)\\
&&\Theta^{(U\,\,U)_L}_{f_1f_2\,\,g_1g_2}=\eta^{U_L}_{f_1f_2}+\eta^{U_L}_{g_1g_2}-
\eta^{U_L}_{f_1+g_1\,f_2+g_2}=\frac{\delta}{4}(2f_1g_1+3f_1g_2+3g_1f_2)\\
&&\Theta^{(G\,\,G)_L}_{f_1f_2\,\,g_1g_2}=\eta^{G_L}_{f_1f_2}+\eta^{G_L}_{g_1g_2}-
\eta^{G_L}_{f_1+g_1\,f_2+g_2}={\delta}(f_1g_2+f_2g_1).
\end{eqnarray}

Qualitative estimates for the  contact exponents  in $d=3$ can be found by direct substitution
of $\varepsilon=1$ in the above relations. To discuss the physical interpretation of these results,
let us consider Fig. \ref{th}, comparing the cases of pure lattice and LR disorder with $a=2.2$ and $a=2.7$.
Fig. \ref{th}a presents the contact exponent $\Theta^{U\,\,U}_{f_1f_2\,\,0g_2}$ governing the interaction
between a star copolymer and a homogeneous star with $g_2$ arms of RWs. We fix $g_2=8$ and change $f_1$ and $f_2$
in such a way that $f_1+f_2=8$.
 The case $f_1=0$, describing two stars of RWs, results in zero value contact exponents and thus the absence of interaction.
Increasing the parameter $f_1$ (SAW component) leads to the gradual increase of the  strength of the interaction.
For $f_1=8$, we have a star of SAWs interacting with a star of RWs with maximal interaction strength.
Fig. \ref{th}b depicts the situation of a star copolymer interacting with a star of $g_1=8$ SAWs.
Again, we change $f_1$ and $f_2$ as above. The case $f_1=0$ describes a star of SAW interacting with
a star of RWs and is a particular case of Fig. \ref{th}a described above.
Increasing  $f_1$  leads to a gradual decrease of the  strength of the interaction.
For $f_1=8$, we have two interacting stars of SAWs with minimal interaction strength.
Fig. \ref{th}c depicts a  situation of two interacting star copolymers with $f_1$, $f_2$ and $g_1$, $g_2$ arms correspondingly.
We fixed $g_1=g_2=4$ and again change $f_1$ and $f_2$ as described above. The case $f_1=0$ describes a copolymer star interacting with
a star of RWs. Increasing the  parameter $f_1$  leads to a gradual increase of the  strength of the interaction,  
until
it reaches its maximal value at $f_1=8$, corresponding to the interaction between a star copolymer and a star of SAWs.
The case $f_1=f_2=4$ describes the interaction between two identical copolymer stars.

Finally, we conclude, that in all situations considered above, the presence of correlated disorder leads to
an increase of the contact exponent. The stronger the correlation (the smaller the value of correlation parameter $a$), the stronger is the interaction between polymers in such an environment. Let us recall, that the exponent $\Theta^{(S\,\,S)_L}_{f_1f_2 \,\,g_1g_2}$ corresponds to the situations of two
interacting homogeneous polymer stars of $f=f_1+f_2$ and $g=g_1+g_2$ arms in solution in the presence of long-range-correlated disorder. This problem has previously been analyzed  \cite{Blavatska06} using a two-loop expansion series  for $\Theta^{(S\,\,S)_L}_{fg}$ in $d=3$.
The quantitative estimates obtained predict a
decrease of the contact exponents with the strength of the disorder correlations in contrast to our present $\varepsilon,\delta$-expansion results. Revising the resummation as performed in \cite{Blavatska06} we 
conclude
that the number of terms in the two-loop expansion is probably too small to rely on those quantitative results. 

\section{Conclusions}

In the present paper, we have studied the scaling properties of copolymer stars,
consisting of $f_1$ arms of polymer species 1 and $f_2$ arms of species 2 in a 
solution  in which one or more of the intra- and interspecies interactions are found to be at their $\Theta$-point 
with the further complication of a disordered environment 
with correlated  structural defects.
We assume that the disorder is correlated with a
power-law decay of the pair correlation function $g(x)\sim
x^{-a}$ at large distance $x$. This type of disorder is known to be relevant
for simple polymer chains \cite{Blavatska01,Blavatska01a} and homogeneous polymer stars \cite{Blavatska06},
and we address  the question of the scaling of copolymer stars in this situation.

Considering the $f_1$-arm  absorbing star polymer with a special trap placed at the center of the star
where $f_2$ free particles (RWs) are to be trapped simultaneously,
the reaction rate of this diffusion-limited reaction is found to scale with exponents, connected to
the spectrum of critical exponents $\eta_{f_1f_2}$ of star copolymers \cite{Cates87}.
Such a process is an example of a so-called diffusion-limited reaction, with the rate depending on the sum of
diffusion coefficients of the reactants.
Another example, where star exponents govern physical behavior concerns the short-range interaction  between the  cores
of star polymers in a good solvent.
The present study aims to analyze the impact of structural disorder in the environment on these processes.

In the frames of the field-theoretical renormalization group approach,
we obtain  estimates for the critical exponents $\eta_{f_1f_2}$ up to the first order of an $\varepsilon=4-d,\delta=4-a$-expansion,
which belong to a new universality class. In particular, this enables us to conclude that the rates of diffusion-limited
reactions are slowed down by the presence of long-range-correlated disorder.
The crucial point is that while long-range correlated disorder
apparently does not influence the RWs and thus the universal behavior of diffusion itself,
the fact that the absorbing polymer changes its conformation in the LR background leads to a change of the 
rate with which  particles are  absorbed (or catalyzed) on specific sites of the polymer.

The contact exponents, governing the repulsive interaction between two star
copolymers in correlated disorder, are found to be larger than in the pure solution case.
The stronger the correlation of the defects, the stronger is the interaction between
polymers in such a disordered environment.

\section*{Acknowledgment}
This work was supported by the Applied Research Fellow Scheme of
Coventry University and by the Austrian Fonds zur
F\"orderung der wissenschaftlichen Forschung under Project
No. P19583-N20. V.B. wishes to thank for a grant of the National Academy of Sciences of Ukraine Committee for young scientists.


\begin{thebibliography}{50}
\bibitem{FP}
Note that some of these solutions are absent because of the special form of the $\beta$-functions. Therefore, we arrive to 27 FPs.
\end{thebibliography}
\end{document}